\documentclass[aps,pra,twocolumn,groupedaddress]{revtex4}
\usepackage{graphics}

\begin{document}

\title{Ontological Models in Quantum Mechanics: What do they tell us?}

\author{L. E. Ballentine}
\email[]{ballenti@sfu.ca}
\affiliation{Dept.~of Physics, Simon Fraser University,
   Burnaby, B.C.~V5A 1S6, Canada}

\date{\today}

\begin{abstract}
The consequences of the theorems about ontological models are studied.
\emph{Maximally $\psi$-epistemic} is shown to be equivalent to the conjunction of
two other conditions, each of which can be realized in Hilbert spaces of arbitrary
dimension, but which cannot occur together for $d>2$. I argue that current theorems 
do not invalidate an epistemic interpretation of quantum states. 
A new condition, called \emph{functionally $\psi$-epistemic}, is
introduced, which, were it to be excluded, would signal the exclusion of epistemic
interpretations.
\end{abstract}


\maketitle


\section{Introduction}

Debates over the status of the \emph{quantum state} concept
have been ongoing since the beginning of the theory.
Is the state $\psi$ a \emph{physically real} object (ontic interpretation), 
or is it an abstract entity that merely provides \emph{information} about 
the system (epistemic interpretation)?
The introduction of \emph{ontological models} by Harrigan and
Spekkens \cite{HS} created a framework whereby the discussion could be made
much more precise. Now the arguments are based on mathematically proven theorems,
in addition to philosophical arguments.
But the growing number of theorems, combined with some peculiar terminology,
has led to confusion. 
We read of \emph{maximally $\psi$-epistemic} models \cite{Mar}\cite{L-Mar}, 
and \emph{maximally nontrivial $\psi$-epistemic} models \cite{ABCL}.
The two terms are not synonymous; moreover, the authors of the first
prove the nonexistence of that class of models, whereas the authors of the
second provide constructive examples of their class.
Several other classes of ontological models have been discussed, many of
which have constructive (though often artificial) examples.
If this was not already complicated enough, it has been claimed to prove \cite{CR1}
that the only \emph{mathematically} consistent interpretation is that
 \emph{the wave function of a system be in one-to-one correspondence with
its elements of reality}.
Yet the observational inadequacy of that interpretation was already well known
to Einstein and Schr\"odinger in 1935, since in that interpretation, Schr\"odinger's
cat would have neither the property of being live nor being dead.
Taken together, all of these claims seem to indicate that no satisfactory
interpretation of quantum theory is even possible!

``A good joke should not be repeated too often,'' as Einstein said to Heisenberg.
In this paper I shall try to overcome the confusion that may come from a casual
reading of the recent literature, with its apparently conflicting claims.
I shall limit myself to ontological models of \emph{single component} systems,
and so will not treat Bell's theorem, nonlocality, or entanglement.
Those topics have their own extensive literature, and including them would make
for much too long a paper.

\section{Ontological models}

In an ontological model, we posit a space $\Lambda$ of \emph{ontic states},
which underlie or supplement the quantum state $\psi$. 
The two primary quantum-mechanical concepts of \emph{state preparation}
and \emph{measurement} are each represented in an ontological model.
A preparation $S_P$ of the quantum state $\psi$ actually yields some ontic
state $\lambda \in \Lambda$. A repetition of the same $S_P$ may yield a
different $\lambda$, and the probability distribution of the resulting ontic 
states, $\mu(\lambda | \psi,S_P)$, 
is called the \emph{epistemic state} associated with $\psi$. It satisfies
\begin{equation} \label{munorm}
 \int_{\Lambda} \mu(\lambda | \psi,S_P) d\lambda = 1\ .
\end{equation}
The possible outcomes of a measurement of the observable $M$  may be labeled by
its eigenvalues, or (as is more convenient here) by its eigenvectors $\{\phi_k\}$.
The probability of obtaining the  $k$'th outcome is given by
the \emph{response function}, $\xi(\phi_k|\lambda,S_M)$, where $S_M$ denotes the
particular measurement method used.
If $\xi$ takes on only the values $0$ or $1$, the model is called
\emph{outcome-deterministic}. If $\xi$ may take on values between 
$0$  and $1$, the model is called \emph{outcome-indeterministic}.

There may be more than one preparation method, $S_P$, that yields the same 
quantum state. If the epistemic state $\mu(\lambda | \psi,S_P)$ depends
nontrivially on $S_P$ for the same $\psi$, the model is called
\emph{preparation contextual}. Similarly, there may be more than one way
to measure $M$. If the response function $\xi(\phi_k|\lambda,S_M)$
depends nontrivially on the method $S_M$, the model is called 
\emph{measurement contextual}. 
When contextuality is not relevant, the notations $S_P$ and $S_M$ may be
omitted for brevity.

The model is required to reproduce the quantum statistics for measurement
outcomes. Without serious loss of generality, we may consider only a
projective observable, $M = |\phi\rangle\langle\phi|$.
Then if the quantum state $\psi$ is prepared, the probability of a positive 
outcome of the measurement will be
\begin{equation} \label{measurement}
 \int_{\Lambda} \xi(\phi|\lambda,S_M)\, \mu(\lambda|\psi,S_P)\, d\lambda
                   = |\langle\phi|\psi\rangle|^2 \ .
\end{equation}

Quantum mechanics has an interesting reciprocal relation between state preparation
and measurement.\\
{\it Quantum Certainty:} A system that is prepared in the state
$\psi$ will \emph{always} pass the test of measuring the
projector $|\psi\rangle\langle\psi|$.\\
{\it Converse:} The state $\psi$ is the \emph{only} state 
that will pass the projective measurement filter $|\psi\rangle\langle\psi|$ with certainty. \\
Since both \emph{Quantum Certainty} and its converse hold in quantum
mechanics, there is a relationship of \emph{preparation--measurement reciprocity},
which will be called \emph{reciprocity}, for short. \\
 
An ontological model must satisfy \emph{Quantum Certainty}, otherwise it would
not agree with a prediction of quantum mechanics.
Hence we must require
\begin{equation} \label{q-cert}
 \int_{\Lambda} \xi(\psi|\lambda,S_M)\, \mu(\lambda|\psi,S_P)\, d\lambda = 1 \ .
\end{equation}
The \emph{supports}, in the ontic state space $\Lambda$, of the two functions 
in the integrand above, are of considerable interest. Define 
$\Lambda_\psi = \mbox{Supp}(\mu(\lambda|\psi))$ as the subset of $\Lambda$ for which 
 $\mu(\lambda|\psi) > 0$.
Similarly, $\mbox{Supp}(\xi(\psi|\lambda))$ is the subset
of $\Lambda$ within which $\xi(\psi|\lambda)$  is nonzero.
We also define the \emph{core support} of the response function, 
 $\mbox{Core}(\xi(\psi|\lambda))$, as the subset of $\Lambda$ within which
 $\xi(\psi|\lambda) = 1$.
The relation between Core and Supp is illustrated in Figure 1.

 \begin{figure}
 \includegraphics{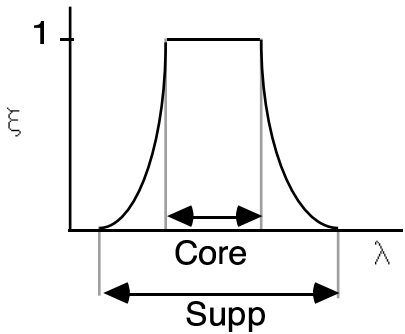}
 \caption{\label{Fig.1} Core support Core$(\xi(\psi|\lambda))$,
    and total support Supp$(\xi(\psi|\lambda))$.}
 \end{figure}

Notice that the range of integration in (\ref{q-cert}) may be reduced from $\Lambda$
to the subset $\Lambda_\psi$, since only there is the integrand nonzero.
But since $\mu(\lambda|\psi)$ is non-negative and satisfies (\ref{munorm}), the requirement
(\ref{q-cert}) can be satisfied only if $\xi(\psi|\lambda) = 1$ for all $\lambda$
inside $\Lambda_\psi$. Hence we conclude that
\begin{equation} \label{mu-core-supp}
  \Lambda_\psi \subseteq {\rm Core}(\xi(\psi|\lambda)) 
               \subseteq {\rm Supp}(\xi(\psi|\lambda))    
\end{equation}

This relation is very useful in clarifying several properties of ontological 
models. The first inclusion ensures that any ontic state that may be produced by
a preparation of $\psi$ will also pass the projective filter $|\psi\rangle\langle\psi|$,
so the model satisfies \emph{Quantum Certainty}.
But the \emph{converse} need not hold. It is possible for the set of ontic states that
pass the filter $|\psi\rangle\langle\psi|$ with certainty to be larger than the
set $\Lambda_\psi$ that can be prepared with the quantum state $\psi$. 
Thus \emph{reciprocity} does not hold for all ontic models.
The condition for \emph{reciprocity} is 
\begin{equation} \label{recip}
 \Lambda_\psi = {\rm Core}(\xi(\psi|\lambda))\ . 
\end{equation}
An \emph{outcome-deterministic} model satisfies
\begin{equation} \label{o-det}
  {\rm Core}(\xi(\psi|\lambda)) = {\rm Supp}(\xi(\psi|\lambda))\ ,
\end{equation}
whereas an \emph{outcome-indeterministic} model satisfies 
\begin{equation} \label{o-indet}
  {\rm Core}(\xi(\psi|\lambda)) \subset {\rm Supp}(\xi(\psi|\lambda))\ .
\end{equation}
The inelegant term \emph{deficiency} has been used \cite{HR}
for the relation 
  $\Lambda_\psi \subset {\rm Supp}(\xi(\psi|\lambda))$.
It is apparent from (\ref{mu-core-supp}) that \emph{deficiency} may alternatively
be defined as 
\begin{equation} \label{defic}
  \mbox{Deficiency} = \mbox{nonreciprocity}\;  \mbox{OR}\; \mbox{outcome-indeterminacy}
\end{equation}

\section{$\psi$-Epistemic Models, Maximal and otherwise}

Harrigan and Spekkens \cite{HS} define two principal classes of ontological
models: \emph{$\psi$-ontic} and \emph{$\psi$-epistemic}. 
Intuitively speaking, \emph{$\psi$-ontic} means that the quantum state $\psi$
is an element of physical reality, whereas \emph{$\psi$-epistemic} means that
$\psi$ is an abstract entity that provides information about the system.

These intuitive notions are supplemented by a more precise mathematical definition.
A model is considered to be \emph{$\psi$-ontic} if the specification of the ontic
state $\lambda$ uniquely determines the quantum state $\psi$.
For this to be true, it is necessary that the preparations of any pair of 
different quantum states, $\psi$ and $\phi$, should yield ontic state distributions
whose supports, $\Lambda_\psi$ and $\Lambda_\phi$, do not overlap.
Two subclasses are identified. 
In a \emph{$\psi$-complete} model, $\psi$ is the \emph{only} ontic state variable. 
In a \emph{$\psi$-supplemented} model, there are some other ontic variables in
addition to $\psi$. 
(One could identify a third class, in which specifying the ontic variable(s) $\lambda$, 
excluding $\psi$, enables $\psi$ to be uniquely calculated. But since no such
models have yet been constructed, the point is moot.)

Although the class \emph{$\psi$-ontic} was carefully defined, the class 
\emph{$\psi$-epistemic} was merely defined as the complement of \emph{$\psi$-ontic}.
This allows trivial $\psi$-epistemic models, such as ones for which  
there is only a single pair of states, $\psi$ and $\phi$, for which $\Lambda_\psi$ 
and $\Lambda_\phi$ overlap.
However, there are nontrivial $\psi$-epistemic models. 
Indeed, Aaronson \emph{et al.} \cite{ABCL} have shown that there are 
\emph{maximally-nontrivial} $\psi$-epistemic models, for which the supports 
$\Lambda_\psi$ and $\Lambda_\phi$ overlap for \emph{every} non-orthogonal 
pair, $\psi$ and $\phi$. These models can be constructed for any finite dimension
of the Hilbert space.

Following this line, the most ambitious goal would be to construct
\emph{maximally-epistemic} models, for which the overlap probability of any
two state vectors, $|\langle\phi|\psi\rangle|^2$, is fully accounted for
by the overlap of the corresponding distributions, $\mu(\lambda|\phi)$ and
$\mu(\lambda|\psi)$. 
A way to implement this condition would be to require that all of the
contributions to the integral on the left side of (\ref{measurement})  
should come from the overlap region $\Lambda_\phi \cap \Lambda_\psi$.
We have the elementary relations
\begin{eqnarray}
 \int_{\Lambda_\phi} \mu(\lambda|\psi)\, d\lambda 
       = \int_{\Lambda_\phi} \xi(\phi|\lambda)\, \mu(\lambda|\psi)\, d\lambda\qquad \nonumber\\
 \qquad\le \int_{\Lambda} \xi(\phi|\lambda)\, \mu(\lambda|\psi)\, d\lambda
             = |\langle\phi|\psi\rangle|^2 .
\end{eqnarray}
(Note: The first line uses $\Lambda_\phi \subseteq {\rm Core}(\xi(\phi|\lambda))$
from (\ref{mu-core-supp}), and the integrals
are effectively over $\Lambda_\phi \cap \Lambda_\psi$.)
A \emph{maximally-epistemic} model is defined to satisfy the above relations
with the inequality replaced by an equality.  If we write
\begin{equation} \label{max-epist}
 \int_{\Lambda_\phi} \mu(\lambda|\psi)\, d\lambda = f(\phi,\psi)\, |\langle\phi|\psi\rangle|^2, 
\end{equation}
where $0 \le f(\phi,\psi) \le 1$, then the condition for a model to be \emph{maximally-epistemic}
is $f(\phi,\psi) = 1$ for all $\phi$ and $\psi$.

The Kochen-Specker model for a spin-$\frac{1}{2}$ system is a nice example
of a \emph{maximally $\psi$-epistemic} model. 
(The original K-S paper predates the framework of ontological models, but both 
\cite{HS} and \cite{HR} give concise descriptions of the K-S model within 
this framework.)
However, Maroney \cite{Mar} has demonstrated that in Hilbert spaces of 3 or more
dimensions, it is impossible to have $f(\phi,\psi) = 1$ for all vectors.
If $f$ is taken to be a constant, he obtains the bound
  $f \le \frac{9}{10}$ for 3 dimensions.
Much tighter bounds have been obtained for higher Hilbert-space dimensions
\cite{Mar}\cite{Leifer}.
Therefore, \emph{maximally $\psi$-epistemic} models do not exist for more
than 2 dimensions. \\

Maroney's theorem \cite{Mar} is surprising, and we should try to better understand
its significance.
To reproduce the statistical predictions of QM, any model must satisfy
\begin{equation} \label{any-model}
 \int_{\Lambda} \xi(\phi|\lambda)\, \mu(\lambda|\psi)\, d\lambda
                = |\langle\phi|\psi\rangle|^2.
\end{equation}
A \emph{maximally $\psi$-epistemic} model must also satisfy 
\begin{equation} \label{max-epist2}
 \int_{\Lambda_\phi} \mu(\lambda|\psi)\, d\lambda
                = |\langle\phi|\psi\rangle|^2.
\end{equation}
These equations must both hold for all choices of $|\psi\rangle$. 
We may assume that any $\lambda\in\Lambda$ can be reached by a preparation of
some $|\psi\rangle$ (else the unpreparable values of $\lambda$ would be 
superfluous, and could be discarded from the model). 
Therefore, it is necessary that $\xi(\phi|\lambda) = 1$ for $\lambda\in\Lambda_\phi$\,,
and that $\xi(\phi|\lambda) = 0$ for $\lambda$ outside of $\Lambda_\phi$.
Hence 
\begin{equation} \label{max-epist3}
  \Lambda_\phi = {\rm Core}(\xi(\phi|\lambda)) 
               = {\rm Supp}(\xi(\phi|\lambda))\, . 
\end{equation}
This equation says that a \emph{maximally $\psi$-epistemic} model must satisfy
\emph{reciprocity} and \emph{outcome-determinacy}.

Because it will be useful in the next section, I state a Corollary of (\ref{max-epist3}):
\begin{eqnarray*}
 \mbox{Maximally $\psi$-epistemic} \Rightarrow \mbox{Measurement noncontextual}.
\end{eqnarray*}
Suppose that $\xi(\phi|\lambda,S_M)$ were to depend nontrivially on the measurement
procedure $S_M$.
We have already shown that $\xi(\phi|\lambda) = 1$ for $\lambda\in\Lambda_\phi$\,,
and is zero elsewhere. 
Since $\Lambda_\phi$ is fixed by the state preparation, it
cannot depend on $S_M$, so there is no way that 
$\xi(\phi|\lambda,S_M)$ can acquire dependence on $S_M$. 
Thus $\xi(\phi|\lambda)$ is actually noncontextual.

Returning to the main argument, the \emph{converse} of (\ref{max-epist3}) is also true. 
This can be seen by following the steps backwards to (\ref{any-model}).
Therefore, we have a new characterization of \emph{maximally $\psi$-epistemic} models:
\begin{eqnarray} \label{max-epist4}
 \mbox{Maximally $\psi$-epistemic} \Longleftrightarrow \mbox{Reciprocity}\qquad\qquad\nonumber \\
                            \mbox{AND}\;\; \mbox{Outcome-determinacy}\quad
\end{eqnarray}
It follows from this implication that any ontological model ($\psi$-ontic included)
can fail to be \emph{maximally $\psi$-epistemic} in only two ways.
It can be \emph{outcome-indeterminate,} or it can fail to be \emph{reciprocal} (the latter
implying that the converse of \emph{Quantum Certainty} does not hold for the model).
In view of Maroney's theorem, every ontological model (except for those that treat only
2-d Hilbert spaces) must be either \emph{outcome-indeterminate} or \emph{nonreciprocal}.

In view of the failure of the attempt to explain the indistinguishability of non-orthogonal
quantum states by the overlap of the corresponding ontic state probability distributions,
we should ask how it is explained within the above two options. 
If the model is \emph{outcome-indeterminate} then the explanation is qualitatively the same
as for ordinary QM without any deeper layer of ontic states.
It is not the case that non-orthogonal quantum states cannot be distinguished --- it is only
that they cannot be distinguished \emph{with certainty}. It is a standard problem of
quantum information theory to devise an optimal measurement for distinguishing
non-orthogonal states. As the two states approach orthogonality, the success rate
approaches 100\%, and as they approach parallelism the success rate approaches that of
pure chance (50\%). But in all cases, the optimal strategy is better than pure chance.
If the ontic states of the model are accessible, then the predictive power of the model
may be greater than that of pure QM. But there will be no qualitative difference.

The next case appears to be quite different. 
In the previous case (\emph{outcome-indeterminate}) we have
\begin{equation} \label{}
  \Lambda_\phi = {\rm Core}(\xi(\phi|\lambda)) 
               \subset {\rm Supp}(\xi(\phi|\lambda))\, , 
\end{equation}
while in this case (\emph{nonreciprocal}), we have
\begin{equation} \label{}
  \Lambda_\phi \subset {\rm Core}(\xi(\phi|\lambda)) 
               = {\rm Supp}(\xi(\phi|\lambda))\, . 
\end{equation}
Yet there is a similarity, in that both cases obey
\begin{equation} \label{defic2}
  \Lambda_\phi \subset {\rm Supp}(\xi(\phi|\lambda))\, . 
\end{equation}
This property that has been burdened \cite{HR} with the undescriptive name \emph{deficiency}.
But since \emph{non-deficiency} is equivalent to \emph{maximally $\psi$-epistemic}, 
perhaps the ugly word \emph{deficiency} can be laid to rest.
Far from being a defect of ontological models, it is actually an aspect of quantum normality.
Example~1 in App.~A (the Beltrametti-Bugajski model) is essentially ordinary QM, without any
additional ontic structures, and it exhibits property (\ref{defic2}).

This section began with a discovery --- the impossibility of
 \emph{maximally $\psi$-epistemic} models in dimensions greater than 2 --- that appeared 
to pose a serious threat to epistemic interpretations of quantum states.
It has concluded by showing that the impossibility of such models is equivalent to
two structural features that affect all ontological models, both $\psi$-epistemic and
$\psi$-ontic. So, thus far, both epistemic and ontic interpretations remain viable.

\section{Contextuality}

The various concepts of \emph{contextuality} have been studied in detail by
Harrigan and Rudolph \cite{HR} and by Leifer and Maroney \cite{L-Mar}.
The latter paper contains the theorem:
\begin{eqnarray} \label{prep-max-KS}
 \mbox{Preparation noncontextual} \Rightarrow \mbox{Maximally $\psi$-epistemic}\nonumber\\
         \Rightarrow \mbox{K-S noncontextual}\qquad
\end{eqnarray}
Both implications are strictly one-directional.
I shall discuss separately the significance of the two parts of this theorem
(for proofs, see \cite{L-Mar}).

\subsection{Preparation Contextuality}

Let $\mu(\lambda|\rho,S_P)$ be the probability of obtaining the ontic state $\lambda$
from a preparation of the (pure or mixed) quantum state $\rho$  by the method $S_P$.
If this probability depends nontrivially on the preparation method $S_P$ for the 
same quantum state $\rho$, this situation is called \emph{preparation contextuality}.
Although this concept could, in principle, be applied to pure states, in practice,
it has only been used for states that are (literally) mixtures.

It is well known that a mixed state operator can be represented as convex sum of
pure components in many ways. For example, the unpolarized state of
an ensemble of spin-$\frac{1}{2}$ particles can be obtained from the following
two mixtures:
\begin{eqnarray} \label{}
  \rho_0 &=& \frac{1}{2} \left(|\uparrow\rangle\langle\uparrow|
                           + |\downarrow\rangle\langle\downarrow|\right) \\
  \rho_0 &=& \frac{1}{2} \left(|\leftarrow\rangle\langle\leftarrow|
                           + |\rightarrow\rangle\langle\rightarrow|\right)
\end{eqnarray}
The first equation describes a mixture of eigenstates of spin $\sigma_z$
(\emph{up} and \emph{down}), while the second describes a mixture of eigenstates 
of spin $\sigma_x$ (\emph{left} and \emph{right}).
The state operator $\rho_0$ is the same in both cases.
But are the distributions of ontic states $\{\lambda\}$ the same in both cases?
For the Kochen-Specker model (Ex.~2 in App.~A), they are not.
The state operator $\rho_0$ has spherical symmetry, but
in the first case, the ontic state distribution has cylindrical symmetry about the
z-axis, while in the second case it has cylindrical symmetry about the x-axis.

As was shown in the previous section, \emph{maximal $\psi$-epistemic} fails
in most cases, therefore, from (\ref{prep-max-KS}), \emph{preparation noncontextuality} must also
fail in those cases. In other words, the ontic state distribution 
$\mu(\lambda|\rho,S_P)$ typically depends, not only on the quantum state $\rho$, 
but also on the particular mixture used to prepare it.

Should this fact be of concern to the supporters of any particular interpretation
of QM? I think not. It is quite common, in both classical and quantum physics,
for a higher level theory to be more
symmetric than the lower level theory that underlies it.
Consider the continuum theory of electrical conductivity,
which is described by the conductivity tensor $\sigma_{\mu\nu}$.
Beneath it lies the atomic theory, in which electrons flow through a crystal
lattice.  The crystal lattice is never isotropic,
but if it possesses 4-fold symmetry, then it is easy to show that this 
implies an isotropic conductivity tensor, $\sigma_{\mu\nu} = \sigma_0 \delta_{\mu\nu}$.
This situation -- that the higher level theory is more symmetric than the lower
level theory beneath it -- is typical.
It is not at all surprising that a similar relation should hold between quantum theory
and the ontological models.

\subsection{Kochen-Specker Theorem}

The theorem of Kochen and Specker \cite{KS} addresses the question:
\begin{quote}
Is it possible that, at any instant of time, the QM observables each possess
a definite value (equal to one of its eigenvalues), regardless of whether
they have been measured? 
\end{quote}
The system that they consider is quite realistic: a particle of spin 1,
and its dynamical variables $S_x^{\,2},\ S_y^{\,2}\ \mbox{and}\ S_z^{\,2}$.
Each of the operators has eigenvalues 0 (nondegenerate) and 1 (doubly degenerate).
These 3 operators are commutative, and their sum is $S_x^{\,2} + S_y^{\,2} + S_z^{\,2} = 2$.
Therefore, in any of their common eigenstates, two of them must have the value
$1$ and one must have the value $0$.
For technical reasons, it is convenient to replace these observables by 
projection operators: $M_x = 1-S_x^{\,2}$\,, and similarly for any other vector 
directions in space.
For any orthogonal triad of directions (x,y,z), the projection
operators satisfy $M_x + M_y + M_z = 1$.

The KS problem is to construct a \emph{valuation} on the set $\{M_i\}$, which
must satisfy the following conditions. \\
 (i) For each vector direction, the value is either 0 or 1. \\
 (ii) For any orthogonal triad of vectors, exactly one of them has the value 1. \\
 (iii) No two orthogonal vectors can both receive the\break value 1. 

Difficulties arise because a vector can belong to more than one triad.
By a heroic effort, Kochen and Specker were able to construct a graph of 117
vectors, for which they proved that no such valuation is possible! \\

\emph{Remark 1:}\ \ The KS theorem is the impossibility of the valuation described above.
There is no mention of \emph{measurement}, and in view of the question posed at the
beginning, there should be none.

There is more than one possible interpretation as to the physical reason behind the
KS theorem.

(PI~1)\ \ The set of observables used in the KS theorem do not have 
predetermined values. Rather, the values emerge as a consequence of the interaction
between the system and the measurement apparatus.

(PI~2)\ \ The outcomes of the measurements are \emph{contextual}.
For example, a measurement of the commuting triple $(M_x, M_y, M_z)$ is one possible
\emph{context} for measurement of $M_x$.
A measurement of $(M_x, M_{y'}, M_{z'})$ is another context.
Here $(y', z')$ are obtained from $(y, z)$ by rotation about the x-axis.
If the measurements are \emph{contextual}, the values obtained for $M_x$ may be different
in these two contexts.

Abbott \emph{et~al.} \cite{ACCS} show that propositions PI~1 and PI~2 are not equivalent.
Each is sufficient to ``explain'' the KS theorem, but neither is necessary.
Note that the KS theorem does not say that \emph{every} observable must be value indefinite
or contextual, but only that it is impossible for the entire set of observables to be
value definite and noncontextual. Abbott \emph{et~al.} show that, for any chosen set of 
observables, it is possible for at least one to be value definite.
An interesting goal of their paper is to identify observables that are
\emph{provably value indefinite,} which could then be used as certifiable quantum 
random-number generators.

\emph{Remark~2:}\ \ The KS theorem applies at the level of QM observables.
Its extension to the level of ontic states is not automatic.

However, the extension is not difficult. Suppose that the response function for the
measurement of $M_i$, $\xi(M_i|\lambda)$, is \emph{outcome deterministic}.
Then, for each ontic state $\lambda$, it will provide a valuation of the observables
$\{M_i\}$ of the kind contemplated in the K-S theorem. But for Hilbert spaces of
dimension 3 (or higher), no such valuation is possible. 
As a consequence of this fact, there are two possibilities. The first is that the
response function be \emph{outcome indeterministic}. It then yields only a probability
for the measurement outcome, and so does not yield a (self-contradictory) 
valuation on the set of observables. The second is that the response function be
\emph{measurement contextual}. This means that $\xi(M_i|\lambda,S_M)$ depends on
the measurement context $S_M$, as well as on the particular observable $M_i$ that is
being measured. This result is worth stating as a theorem, which I shall call the
``KS-OM Theorem'' (KS theorem extended to Ontological Models):\\

{\bf KS--OM Theorem}:\ \  For a quantum system of Hilbert-space dimension 3 or greater,
any underlying ontological model must be \ \emph{outcome indeterministic}\ 
OR\ \ \emph{measurement contextual.}\\

Notice that these two options for ontological models closely parallel the two options
at the QM observable level, PI~1 and PI~2, that were paraphrased from \cite{ACCS}.
But the necessary arguments in \cite{ACCS} were very subtle and complex, whereas
the derivation of the KS-OM theorem was very simple. This confirms, once again,
the utility of the ontological model as a method of investigation.\\

Let us now return to the second implication of the theorem (\ref{prep-max-KS}),
which by logical contraposition becomes\\
 $\mbox{\quad K-S contextual} \Rightarrow \mbox{NOT Maximally $\psi$-epistemic}$.\\ 
The authors did not distinguish between the KS theorem and the KS-OM theorem (naturally
enough, since the latter had not yet been formalized),
and so the exact meaning of their implication above is not quite clear.
Their definition of ``the Kochen-Specker theorem'' in \cite{L-Mar} does not
reflect the conditions of the original KS theorem; however, it is very close to those
of the KS-OM theorem. From its context, their term ``K-S contextual'' seems
to mean the second option of the KS-OM theorem. 
However, a stronger statement is possible.

\emph{Both options} of the KS-OM theorem imply that \emph{Maximally $\psi$-epistemic} 
models are excluded.

From (\ref{max-epist4}), it is evident that the first option,
 \emph{outcome indeterminacy}, excludes \emph{Maximally $\psi$-epistemic} models.
Immediately following (\ref{max-epist3}) was stated a Corollary, to the effect that
a \emph{Maximally $\psi$-epistemic} model must be \emph{measurement noncontextual}.
So the second option of the KS-OM theorem also excludes 
\emph{Maximally $\psi$-epistemic} models.

This was actually proven in \cite{L-Mar} (their \emph{Theorem 1}),
and any confusion is due only to the ambiguity of the term ``KS contextual''.

\section{$\Psi$-Ontic-Complete}

Colbeck and Renner \cite{CR1} have presented a theorem, from which they conclude that
\emph{``A system's wave function is in one-to-one correspondence with its
       elements of reality.''}
Their argument is based on the conclusion of an earlier paper \cite{CR2}, in which they
claimed that 
\emph{``No extension of quantum theory can have improved predictive power.''}
In the context of ontological models, their claims are equivalent to excluding
all models except \emph{$\psi$-ontic-complete}.

The audacity of their claims invites skepticism.
Although their papers are heavily formal, the authors are very helpful 
in pointing out that their conclusions depend on a
certain assumption relating to \emph{Freedom of Choice}. They point out that
several counter-examples to their claims violate that assumption.

Colbeck and Renner represent \emph{free choice} as a random variable that
is uncorrelated with anything in its past light-cone.
The unsatisfactory nature of that characterization is apparent.
For example, democratic elections are based on the \emph{secret ballot},
the purpose of which is to ensure that the voters may excersize \emph{free choice}. 
The voting patterns of most voters are strongly correlated with past events.
Many people vote consistently for the same party, but they may depart from that
pattern for any of several reasons: recent performance of their old party in government,
personal characteristics of particular candidates, self-interest, etc.
But those correlations with events in the past light-cone notwithstanding,
\emph{they are still excersizing free choice!} 

The characterization of \emph{free choice} that Colbeck and Renner use is
unrealistic and unsatisfactory. Since, by their own admission, their results
depend critically on that assumption, it follows that their radical claims 
are null and void. 

Ghirardi and Romano \cite{GR}, in opposition to the claim of \cite{CR2}, have
shown how to construct an ontological model whose predictive power is different,
and possibly greater, than that of quantum theory. The essence of their model
is to separate the ontic state $\lambda$ into two parts $(\nu, \tau)$, where
$\nu$ is inaccessible and must be averaged over, but $\tau$ is accessible.
Let $\hat{A}(a)$ denote a quantum mechanical observable that depends on the
parameter $a$. (A component of spin, $\hat{A}(a) = \vec{\sigma} \cdot \vec{a}$, is an example.)
The average of this observable can be calculated in two steps. 
First average over the inaccessible ontic variable $\nu$, to obtain the 
conditional average, $\langle \hat{A}(a)\rangle_{\psi,\tau}\,$.
Then average over the accessible variable $\tau$ to obtain the usual quantum state
average, $\langle \hat{A}(a)\rangle_{\psi} = \langle\psi|\hat{A}(a)|\psi\rangle$.
The model can be constructed so that the conditional average 
$\langle \hat{A}(a)\rangle_{\psi,\tau}$ is more informative than is the quantum
state average.

Many thought experiments (such as that in Bell's theorem) treat the selection of
the observable to be measured as a \emph{free choice} by the experimenter.
In \cite{GR} \emph{free choice} is represented by allowing the parameter $a$ in
$\hat{A}(a)$ to be arbitrary, and then proving the desired results for all values
of $a$. Thus the manner in which the experimenter chooses  $a$ is
irrelevant. This method is more satisfactory than to speculate about the random
or non-random nature of \emph{free will} \cite{freewill}.

The interpretation of QM in which
\emph{``A system's wave function is in one-to-one correspondence with its
       elements of reality''}
has been criticized, and usually rejected, because of Schr\"odinger's Cat Paradox (1935).
It follows from that interpretation, that the cat possesses neither the attribute of
being \emph{alive} nor the attribute of being \emph{dead}.
If that conclusion is rejected on observational grounds, then the 
ontological model \emph{$\psi$-complete} will not lead to a satisfactory interpretation
of quantum theory. However, \emph{$\psi$-epistemic} models and 
\emph{$\psi$-ontic-supplemented} models remain as viable candidates.
In all of those models, the cat may be either \emph{alive} or \emph{dead}, but the 
quantum state does not provide us with the information as to which is the case.

\section{Conclusions}

The broadest consequence of introducing ontological models into 
the foundations of QM, is a sharpening of the discussions. 
A woefully common feature, in the past, was that each protagonist had some interpretation
of the quantum state in mind, but never stated clearly what it was,
and so might shift between \emph{ontic} and \emph{epistemic} 
interpretations whenever he found it convenient.
The introduction of ontological models has made people aware of the distinction between 
\emph{ontic} and \emph{epistemic}.

But beyond this raising of awareness, it has provided a mathematical framework,
within which various possibilities can be evaluated. 
Ontic and epistemic models were distinguished by the overlap of the
\emph{epistemic states} (distributions of ontic states) that are associated with
the preparations of different quantum states.
This overlap varies from zero for \emph{$\psi$-ontic} models, through various degrees
of overlap for \emph{$\psi$-epistemic} models, to the limit of 
\emph{maximally $\psi$-epistemic}. 

It came as a surprize when Maroney's theorem \cite{Mar} demonstrated that 
\emph{maximally $\psi$-epistemic} models are not possible for Hilbert spaces of
more than 2 dimensions. Indeed, some people concluded (prematurely) that epistemic
interpretations of quantum states had been ruled out.
However, it was shown earlier in this paper that \emph{maximally $\psi$-epistemic} is
logically equivalent to the conjunction of \emph{preparation-measurement-reciprocity}
AND \emph{outcome-determinism}. Thus any model (\emph{$\psi$-epistemic} or
\emph{$\psi$-ontic}) can fail to be \emph{maximally $\psi$-epistemic} in only two ways:
either through \emph{nonreciprocity} or by \emph{indeterminism}.
It had been hoped that the inability to distinguish non-orthogonal quantum states
with certainty might be explained through the overlap of the corresponding 
ontic state distributions. That proved to be impossible, but \emph{measurement-outcome 
indeterminism} can also account for non-distinguishability of non-orthogonal quantum states,
and this works for both \emph{$\psi$-epistemic} and \emph{$\psi$-ontic} models. \\

An ontological model has two essential parts: the distribution $\mu(\lambda|\psi,S_P)$
that characterizes quantum state preparation, and the response function 
$\xi(M_i|\lambda,S_M)$ that characterizes measurement. 
The most important structure of the model is the separation of \emph{preparation}
from \emph{measurement}, with information passing only via the ontic state variables.
If the state $\psi$ has a direct effect on the measurement outcome, then $\psi$ should
be classified as an \emph{ontic} variable.
The definition in \cite{HS} of \emph{$\psi$-ontic} and \emph{$\psi$-epistemic} in terms 
of the overlap between the distributions of ontic states, 
deals only with \emph{preparation}, and does not deal with the passing of information 
to the measurement.
Therefore, I shall introduce a new definition for this purpose.

{\bf Def\/ine} a model to be \emph{functionally $\psi$-epistemic} if the measurement
response function $\xi(M_i|\lambda,S_M)$ does not depend on the state $\psi$.
Otherwise, if the response function $\xi(M_i|\psi,\lambda,S_M)$ depends essentially on
$\psi$, the model is \emph{functionally $\psi$-ontic}.

The Kochen-Specker model (Ex.~2 in App.~A) is an example of a 
\emph{functionally $\psi$-epistemic} model, but it is restricted to Hilbert spaces
of 2 dimensions.

Under the old definition in \cite{HS}, we could convert a \emph{$\psi$-epistemic} model
into a \emph{$\psi$-ontic-supplemented} model by declaring -- verbally
-- that $\psi$ is to be considered \emph{ontic}. That would be quite artificial,
but it is allowed by that definition. 
The new definition of \emph{functionally $\psi$-epistemic} does not allow such
a loophole.

I claim that the essential feature of an \emph{epistemic} interpretation of quantum
states is better captured by the concept of \emph{functionally $\psi$-epistemic} 
than by the definition of \emph{$\psi$-epistemic} in \cite{HS}.
The nonexistence of \emph{maximally $\psi$-epistemic} models (for $d \ge 3$) does not
rule out \emph{epistemic} interpretations. However, if \emph{functionally $\psi$-epistemic}
models should turn out to be excluded, then \emph{epistemic} interpretations would, indeed,
be excluded with them.

\section*{Appendix A -- Some Ontological Models}

Some examples of Ontological Models are described in Table I.
The last 3 columns indicate whether the model has the attributes of
\emph{Preparation-measurement-reciprocity, Outcome-determinism, \emph{and}
Measurement-contextuality}.
Recall that all models satisfy (Eq.~\ref{mu-core-supp}),\\
\ \  $\Lambda_\psi \subseteq {\rm Core}(\xi(\psi|\lambda))
               \subseteq {\rm Supp}(\xi(\psi|\lambda))$, and that\\
\emph{Reciprocity} means $\Lambda_\psi = {\rm Core}(\xi(\psi|\lambda))$, and \\
\emph{Determinism} means ${\rm Core}(\xi(\psi|\lambda)) = {\rm Supp}(\xi(\psi|\lambda))$.\\

\begin{table}[h]
\caption{Examples of Ontological Models}
\begin{tabular}[t]{r||c|p{10ex}|c|c|c}
& Name & Type & Reciprocity & Determinism & Contextual\\ \hline
1 & B--B & ontic-complete  & yes & no & no \\
2 & K--S & epistemic (d=2) & yes & yes & no \\
3 & Aaronson & ontic-supplem. & yes & no & yes \\
4 & Bell 1st & ontic-supplem. & no & yes & yes \\
5 & Bell 2nd & ontic-supplem. (d=2)& no & yes & no \\
6 & Aerts & ontic-complete (d=2) & yes & no & no \\
7 & W--S & ontic-supplem. & no & yes & yes \\
\end{tabular}
\end{table}

Notice that, of the properties in the last 3 columns, not all combinations 
of \emph{yes} and \emph{no} are represented.
Since \emph{reciprocity} AND \emph{determinism} imply that the model is
\emph{maximally $\psi$-epistemic} (\ref{max-epist4}), and the Corollary of
(\ref{max-epist3}) requires \emph{maximally $\psi$-epistemic} to be 
\emph{measurement-noncontextual}, it follows that the combination
\emph{yes--yes--yes} is impossible.
It remains to be determined whether models exhibiting \emph{no--no--yes}
and \emph{no--no--no} are possible.
It would be particularly interesting to exhibit a \emph{measurement-noncontextual}
model for $d>2$, since the KS-OM theorem seems to permit it, provided the model
is \emph{outcome indeterministic}.

Models 1--6 in Table I are discussed in detail in \cite{HR}, and in those cases 
where I have nothing to add, I shall merely refer to that paper.
For original references, see \cite{HR}.

1. \emph{Beltrametti-Bugajski model:}\ \ This model is really just ordinary QM, written
in the language of ontological models.  The ontic state space $\Lambda$ is the 
projective Hilbert space of quantum states, and
 $\mu(\lambda|\psi)\,d\lambda = \delta(\lambda - \psi)\,d\lambda$.
The response function for a projective measurement of $|\psi\rangle\langle\psi|$ is
 $\xi(\psi|\lambda) = |\langle\lambda|\psi\rangle|^2$.
The core-support of $\xi(\psi|\lambda)$ is the single point $\lambda = \psi$,
whereas its full support consists of $\Lambda$ minus the set of measure zero for
which $\langle\lambda|\psi\rangle = 0$. Thus the model satisfies \\
$\Lambda_\psi = {\rm Core}(\xi(\psi|\lambda))
               \subset {\rm Supp}(\xi(\psi|\lambda))$,
showing that it satisfies \emph{preparation-measurement-reciprocity},
but is ``deficient.''

 2. \emph{Kochen-Specker model:}\ \ See \cite{HR} for details.
The K-S model for a spin-$\frac{1}{2}$ system exhibits everything that is ``nice'' 
in ontological models:
\emph{reciprocity, maximally $\psi$-epistemic, functionally $\psi$-epistemic,
outcome-deterministic, \emph{and} measurement-noncontextual}.
It also provides a simple example of \emph{preparation-contextuality}.
Unfortunately, there are fundamental reasons why this combination of nice features 
cannot all be realized for higher dimensional Hilbert spaces.

3. \emph{Aaronson's model:}\ \ See \cite{HR} for details.
This model is similar to Model 1, but it adds a privileged basis, and so becomes
\emph{$\psi$-supplemented.} It was invented for computation-theoretic purposes.

4. \emph{Bell's first model:}\ \ See \cite{HR} for details.

5. \emph{Bell's second model:}\ \ See \cite{HR} for details.
Both of Bell's models supplement the quantum state space with an additional space
of ontic variables. The second model is specialized to spin-$\frac{1}{2}$.

6. \emph{Aerts model:}\ \ See \cite{HR} for details.
As in Model 1, the ontic state of the system is identical with its quantum state.
But there are hidden ontic variables in the measurement apparatus.
If their values were known, the measurement would be \emph{outcome deterministic}.
But since they are unknown, they must be averaged over, leading to an effective
\emph{outcome indeterministic} response function.

7. \emph{Wiener-Siegel model:}\ \ This model is treated in detail by Belinfante
\cite{belin}, (the main point is on p.135).
The ontic variables are the state vector $|\psi\rangle$ and another vector $|\lambda\rangle$.
The latter is unnormalized, and has a Gaussian distribution over Hilbert space.
Let $B = \{|j\rangle\}_{j=1}^N$ be a basis for a projective measurement.
Introduce $a_j = \langle j|\psi\rangle$ and $b_j = \langle j|\lambda\rangle$.
Then the \emph{response function} is defined to be $\xi(i|\psi,\lambda,B) = 1$ if
the ratio $|a_i/b_i|$ is the largest of the set $\{|a_j/b_j|\}$, and is zero otherwise.
The model is \emph{outcome deterministic}, and so must violate \emph{reciprocity}
(since it cannot be \emph{maximally $\psi$-epistemic}).
That it is \emph{measurement contextual} can be seen by keeping the basis vector
$|i\rangle$ fixed and doing a unitary transformation on the remaining set
 $\{|j\rangle\}$ for $j\ne i$. This transformation can change the
identity of the largest of the ratios $\{|a_j/b_j|\}$.

\section*{Appendix B -- Interpretations of the Quantum State}

There is a tendancy to say that the quantum state is either physical real
or it represents knowledge.
In fact the range of options is much greater.
The three dichotomies shown in the rows of Fig.~2: \emph{individual} versus \emph{ensemble},
\emph{ontic} versus \emph{epistemic}, and \emph{objective} versus \emph{subjective},
are not equivalent.
 \begin{figure}[h]
 \includegraphics{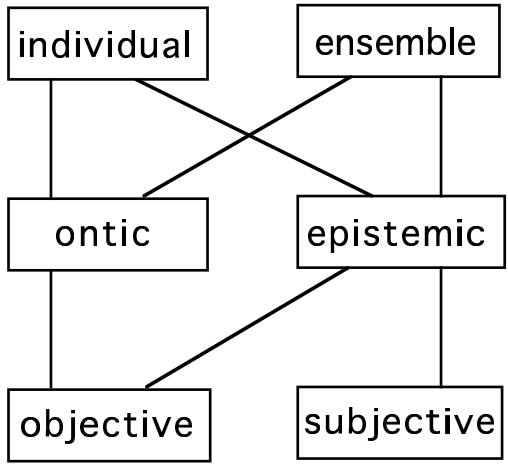}
 \caption{\label{Fig.2} Relations among classes of interpretations of the quantum state.
          The lines between boxes indicate overlap of membership.}
 \end{figure}
A specific interpretation will involve a choice of one class in each row.
(For the record, my own writings on this subject are firmly in the classes of
\emph{ensemble} and \emph{objective}. So far, I maintain an open mind regarding 
\emph{ontic} versus \emph{epistemic}.)

The classes of \emph{ontic} and \emph{epistemic} are central to this subject, 
but the meanings of those terms are not quite the same as those in a dictionary.  
In the commonly used definitions of \cite{HS}, the class \emph{$\psi$-epistemic}
includes everything not included in \emph{$\psi$-ontic}. That being so, we must broaden
the meaning of \emph{epistemic} from its dictionary meaning, pertaining to \emph{knowledge},
to include \emph{information} in general.
Since the fundamental work of Shannon, it has been clear that \emph{information} need not
be subjective. In his theory, a string of bits has the same information content, regardless
of whether it is part of anyone's knowledge, and regardless of whether it contains a
humanly interesting message. Therefore, I have been careful to describe epistemic interpretations
of $\psi$ as carrying \emph{information} about the system, rather than knowledge.
The word \emph{knowledge} begs the question of \emph{Whose knowledge?}, whereas the
word \emph{information} does not.

Moving to the top row of Fig.~2, one can naturaly regard  
an \emph{ensemble} interpretation as epistemic.
The quantum state does not predict events, but only the probabilities of events. 
Hence it provides information about the system, but does not completely describe it. 
But the case of \emph{$\psi$-ontic-supplemented} also yields an ensemble description,
so an ensemble interpretation can be either epistemic or ontic.

Similarly, an \emph{individual} interpretation can be naturally regarded as ontic,
but it is at least logically possible for $\psi$ to give information about
the individual, without itself being an ontic element of reality. I do not know of
an actual example of such an interpretation, so I will give an analogy.
In classical thermodynamics the variables P,V,U,S, and T are regarded as onticly real 
properties of the individual system. 
The partition function of statistical mechanics is not regarded as ontic, but it can
be used to calculate those ontic properties.
It seems logically possible for $\psi$ to relate to the individual quantum system,
in the same way that the partition function relates to the individual thermodynamic
system.

On the bottom row of Fig.~2, it is clear that a \emph{subjective} interpretation
must be epistemic. But an \emph{epistemic} interpretation may be either 
objective or subjective, depending on whether the information that its provides
is objective (independent of any agent) or subjective (knowledge possessed by 
some particular agent).


\begin{thebibliography}{99}
\bibitem{freewill} I strongly advise theoretical physicists to stop talking
about ``consciousness'' or ``free will'', at least until they familiarize
themselves with the subject of Cognitive Science, within which those
concepts are seriously studied.
\end{thebibliography}

\end{document}